\pdfoutput=1

\documentclass[11pt]{article}
\usepackage{booktabs}
\usepackage[preprint]{acl}

\usepackage{times}
\usepackage{latexsym}
\usepackage{graphicx}
\usepackage{subcaption}
\usepackage[T1]{fontenc}

\usepackage[utf8]{inputenc}

\usepackage{microtype}

\usepackage{inconsolata}

\usepackage{graphicx}
\usepackage{enumitem}
\usepackage{listings}
\usepackage{listings}
\usepackage{xcolor}
\usepackage{makecell}
\usepackage{flushend}
\usepackage{balance}

\lstdefinelanguage{ocaml}{
  morekeywords={
    and, as, assert, asr, begin, class, constraint, do, done, downto, else,
    end, exception, external, false, for, fun, function, functor, if, in,
    include, inherit, initializer, land, lazy, let, lor, lsl, lsr, lxor, match,
    method, mod, module, mutable, new, object, of, open, or, private, rec,
    sig, struct, then, to, true, try, type, val, virtual, when, while, with
  },
  sensitive=true,
  morecomment=[s]{(*}{*)},
  morestring=[b]",
}
\lstdefinelanguage{haskell}{
  morekeywords={
    case, class, data, default, deriving, do, else, if, import, in, infix,
    infixl, infixr, instance, let, module, newtype, of, then, type, where, 
    forall, foreign, hiding, qualified, as
  },
  sensitive=true,
  morecomment=[l]--,
  morecomment=[s]{\{-}{-\}},
  morestring=[b]",
  literate={+}{{$+$}}1 {/}{{$/$}}1 {*}{{$*$}}1 {=}{{$=$}}1
           {>}{{$>$}}1 {<}{{$<$}}1 {\\}{{$\lambda$}}1
           {->}{{$\rightarrow$}}2 {<-}{{$\leftarrow$}}2
           {=>}{{$\Rightarrow$}}2
           {>>}{{>>}}2 {>>=}{{>>=}}3
           {|}{{$\mid$}}1
}

\usepackage{listings}
\usepackage{xcolor}

\usepackage{multirow}
\title{\textit{Perish or Flourish?} A Holistic Evaluation of Large Language Models for Code Generation in Functional Programming}

\author{Nguyet-Anh H. Lang$^{1}$, Eric Lang$^{2}$, Thanh Le-Cong$^{2}$\thanks{Corresponding Author: thanhlc@ieee.org}, Bach Le$^{2}$, Quyet-Thang Huynh$^{2}$\\
  $^{1}$ Hanoi University of Science and Technology\\
  $^{2}$The University of Melbourne\\}

\begin{document}
\maketitle
\begin{abstract}

Functional programming provides strong foundations for developing reliable and secure software systems, yet its adoption remains not widespread due to the steep learning curve. Recent advances in Large Language Models (LLMs) for code generation present new opportunities to lower these barriers. However, extensive evaluations of LLMs largely focus on imperative programming languages, and their capabilities in functional programming languages (FP) remain underexplored. To address this gap, we introduce FPEval, a holistic evaluation framework built on FPBench, a new benchmark of 721 programming tasks across three difficulty levels on three mainstream FP languages: Haskell, Ocaml and Scala. FPEval provides compehensive evaluation infrastructures with both test validations with comprehensive test suites and static analysis tools to assess both functional correctness and code style and maintainability. Using this framework, we evaluate state-of-the-art LLMs, including GPT-3.5, GPT-4o, and GPT-5, for code generation in functional programming languages and Java as an imperative baseline. Our results demonstrate that LLM performance in functional programming improves substantially with model advancement; however, error rates remain significantly higher in purely functional languages (Haskell and OCaml) than in hybrid (Scala) or imperative (Java) languages. Moreover, LLMs frequently generate non-idiomatic functional code that follows imperative patterns, raising concerns about code style and long-term maintainability. Finally, we showed that LLMs can partially self-repair both correctness and quality issues when provided with static analysis feedback and hand-crafted instructions for common types of issues.
\end{abstract}

\section{Introduction}

Functional programming is an emerging declarative programming paradigm that conceptualizes computation as the evaluation of mathematical functions rather than a sequence of state-mutating commands in widely-used imperative programming. This model of computation offers distinct advantages over imperative programming, resulting in software systems that are modular, deterministic, and susceptible to formal reasoning. For example, functional programming offers immutability, which ensures that a variable's value cannot be changed during its existence. Immutability enables functional programming to mitigate common sources of software bugs in imperative programming such as side effects and race conditions. Consequently, functional programming is increasingly recognized as a future paradigm for reliable and secure software development~\cite{Achten2013WhyFP, Hughes1989WhyFP, Ray2014ALS}. Despite these benefits, functional programming poses a steep learning curve. The paradigm requires a fundamental shift from imperative commands with mutable state to high-level abstractions of program behaviors such as recursion, higher-order functions, and monads, which many developers find difficult. Consequently, functional programming remains unduly less popular in practice than imperative programming.

The recent rise of coding assistants powered by Large Language Models (LLMs), such as GitHub Copilot~\cite{chen2021codex}, Cursor~\cite{cursor2023}, and Claude Code~\cite{anthropic2025claude}, presents new opportunities for functional programming. By supporting software developers in tasks such as code generation~\cite{2022AlphaCode, chen2021codex}, bug fixing~\cite{10172803, LeCong2026MemoryEfficient}, and technical question answering~\cite{Xu2023AreWR}, these tools offer particular benefits to novice programmers and have the potential to reduce the learning curve associated with programming in general~\cite{Prather2023ItsWT, Kazemitabaar2023StudyingTE}. If similar benefits extend to functional programming, these LLM-based coding assistants could contribute to increasing the adoption of the programming paradigm in practice. Unfortuntately, existing studies on LLM-based coding~\cite{Fan2023LLMSurvey} mainly focus on imperative programming languages, such as Python and Java. In contrast, the applicability of LLMs to functional programming languages, such as Haskell, OCaml, and Scala, remains significantly under-explored.

In this work, we conduct the first comprehensive empirical study to evaluate the capabilities of LLMs for code generation in functional programming. Our objective is three-fold: (1) to assess the ability of LLMs for generating correct and high-quality code in functional programming; (2) to evaluate the quality of LLM-generated code beyond functional correctness, with particular attention to coding style and maintainability; and (3) to evaluate the effectiveness of LLMs on self-repairing their mistakes. To this end, we structure our study using the following research questions:

\begin{itemize}
    \item \textbf{RQ1:} How effective are LLMs for code generation in functional programming languages?
    \item \textbf{RQ2:} What are common coding style and maintainability issues in code generated by LLMs in functional programming languages?
    \item \textbf{RQ3:} How effective are self-repair mechanisms in improving the correctness and code style and maintainability of LLM-generated functional code?
\end{itemize}

To answer these questions, we introduce FPEval, a holistic evaluation framework designed for code generation in functional programming. FPEval is powered by FPBench, a new multi-language benchmark consisting of 721 programming tasks distributed across three difficulty levels (184 easy, 346 medium, and 191 hard) and three languages: Scala, Haskell, and OCaml. To ensure a rigorous assessment on LLM-generated code, we construct a comprehensive test suite for each task, augmenting public test cases from LeetCode with private test cases, which target specific boundary conditions and edge cases for each programmming task. Furthermore, FPEval also integrates well-known static analysis tools, including HLint~\cite{mitchell2024hlint}, OCamlFormat~\cite{ocamlformat2024}, and Scalastyle~\cite{scalastyle2024}, enabling a dual assessment of both the functional correctness and code style and maintainability of the generated code.

Using FPEval, we evaluate state-of-the-art LLMs (GPT-3.5, GPT-4o, and GPT-5) on functional programming tasks, with Java included as an imperative baseline. Our evaluations show that LLM performance on functional programming improves substantially with model advancement, with GPT-5 achieving an approximately 3x increase in the number of functionally-correct code over GPT-3.5. However, a consistent performance gap persists between purely functional languages (Haskell and OCaml) and hybrid or imperative languages (Scala and Java).
Beyond correctness, we find that a significant portion of LLM-generated code has poor style and maintainability. In particular, LLMs frequently produce non-idiomatic functional code that follows imperative patterns rather than best practices in functional programming. Notably, the prevalence of such low-quality code increases alongside gains in functional correctness, indicating a form of reward hacking in which models optimize for correctness while neglecting non-functional properties. These findings highlight the need for future LLM training and evaluation to explicitly incorporate functional programming best practices. Finally, we show that LLMs can partially self-repair both correctness and quality issues when provided with static analysis feedback and hand-crafted instructions for common types of issues.

In summary, our main contributions include:

\begin{itemize}

\item We introduce FPEval, a holistic framework for evaluating LLM code generation in the functional programming paradigm. FPEval integrates FPBench, a curated dataset of 721 programming tasks with executable test suites in Scala, Haskell, and OCaml, and an automated evaluation pipeline that leverages software testing and static analysis tools to rigorously assess both the correctness and the code style and maintainability of generated code. 

\item We present the first empirical evaluation of LLMs for functional programming code generation, systematically assessing the functional correctness and code style and maintainability of generated code, and the LLMs’ capacity for self-repair.

\item We provide a detailed analysis of code style and maintainability issues in LLM-generated functional code and demonstrate the potential LLM's self-repair for improving the code style and maintainability issues with static analysis feedbacks and detailed instructions for common issues. 

\item We advance research in code generation for functional programming by releasing FPEval under Apache 2.0 License, 
lowering barriers for academic research and establishing foundational benchmarks and metrics for future work. 
An early access version of FPEval can be found at~\url{https://github.com/thanhlecongg/FPEval}.

\end{itemize}

\section{Benchmark Construction and Evaluation Pipeline}

\subsection{Benchmark Construction}

\begin{figure}[t]
\centering
\includegraphics[width=0.8\linewidth]{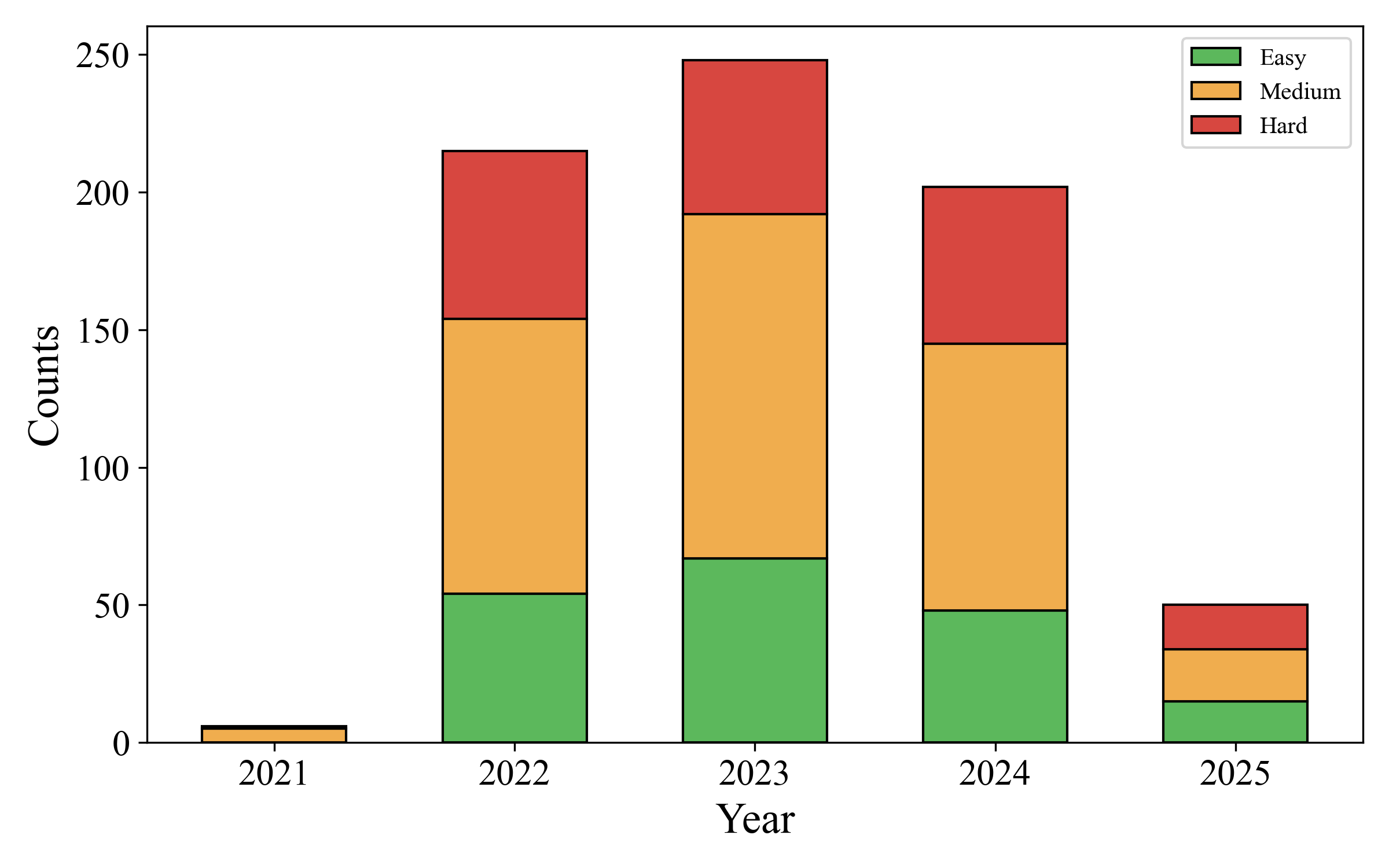}
\caption{Task distribution across time}
\label{fig:distribution_time}
\end{figure}

\textbf{Data Collection}. To support a rigorous empirical study in the code generation of LLMs in functional programming, we curated a comprehensive benchmark dataset from LeetCode~\cite{leetcode}, a widely-used platform for competitive programming and technical interviews. From the full collection, we filtered out premium-only content to select 721 publicly available algorithmic tasks. These tasks span three distinct difficulty levels consisting of 184 easy, 346 medium, and 191 hard programming tasks. For each task, we extracted the problem description, I/O constraints, and public test cases via LeetCode’s public API and custom HTML parsers. Collected public test cases were incorporated into prompt construction and later reused for initial correctness checks. Figures~\ref{fig:distribution_time} illustrate the diversity of the dataset, i.e., FPBench. Our data collection period from 2021 to 2025 enables the capture of evolving trends in algorithm design and problem types. FPBench also include programming tasks with diverse difficulty levels, enabling us to systematically evaluate model performance under diverse problem complexity. 

\paragraph{Language Template Generation.}
A critical challenge to evaluating functional programming on LeetCode tasks is the lack of support for widely-used functional programming languages such as Haskell and OCaml. Specifically, LeetCode do not provide any templates for these functional programming, yielding difficulties for LLMs to understand I/O and typing requirements. To bridge this gap, we implemented a transpiler that constructs syntactically and semantically valid starter templates for Haskell and OCaml from Python templates provided by LeetCode. This translastion was challenging due to fundamental differences in programming paradigms and type system, e.g., dynamic versus static typing, imperative versus functional control flows, and the lack of one-to-one mappings for many language constructs. Our transpiler addresses these discrepancies by (1) performing structural translation of Python function definitions to typed signatures in Haskell and OCaml, including parameter binding and return types; (2) inferring types for nested and compound data structures (e.g., mapping \texttt{List[List[int]]} to \texttt{[[Int]]}); (3) normalizing I/O semantics to conform to functional patterns while preserving the semantics of the original task specification. 

\subsection{Evaluation Pipeline} 

\paragraph{Test validation.} The FPEval pipeline begins by assessing the functional correctness of the generated code. While prior works~\cite{10.1145/3643674} typically utilize public test cases from LeetCode, this test suite is often limited in coverage and may fail to detect subtle logical errors. To mitigate this, we augment each task with a suite of \textit{private test cases}. Following Huang et al.~\cite{lcb2024}, we employ GPT-4o to synthesize these rigorous test cases directly from task descriptions to target boundary conditions, corner cases, and degenerate inputs (see Appendix~\ref{appendix:prompt_testcases} for the specific prompts). Note that, executing these tests within the functional paradigm also presents a critical challenge. Unlike Python, which supports flexible and dynamic execution, statically typed languages like Haskell and OCaml require rigorous type scaffolding, strict compilation, and runtime isolation. To address this, we developed a specialized test infrastructure for functional programming languages using test templates with the standard library and a Docker-based isolated runtime environment, as detailed in Appendix~\ref{appendix:test_execution}.

\paragraph{Code Style and Maintainability Evaluation.} Beyond functional correctness, the code style and maintainability of the generated code is a critical dimension of our evaluation. A solution may satisfy all test cases while still being low quality with respects to best practices in functional programming. Such code accumulates technical debt, rendering systems difficult to maintain~\cite{TechnicalDebt} and even potentially introducing latent safety issues~\cite{Perry_2023}. To systematically quantify this dimension, FPEval integrates widely-used static analysis tools that check weather a code snippet strictly adhere functional programming best practices. Specifically, we ultilize \texttt{HLint} and \texttt{GHC} for Haskell, \texttt{Dune} and \texttt{OCamlFormat} for Ocaml, and \texttt{Scalastyle} for Scala. A example categorization of the rules applied by these tools is provided in Appendix~\ref{appendix:static_analysis_rules}.

\subsection{LLM Selection and Configurations}

To ensure a comprehensive evaluation across varying levels of capability and cost, we evaluate three Large Language Models of OpenAI GPT series including GPT-3.5-turbo~\cite{OpenAI2022ChatGPT}, GPT-4o~\cite{openai2024gpt4technicalreport} and GPT5~\cite{OpenAI2025GPT5}. We adopt a \textit{zero-shot prompting} strategy, in which the model is provided solely with the natural language task description and the language-specific starter template(see Listing~\ref{code:prompt} for an illustration). 
We configure the generation parameters for GPT-3.5-turbo and GPT-4o with \texttt{temperature = 0.7}. For GPT5, we utilize its default  \texttt{temperature = 1.0}, as this model architecture fixes the temperature to optimize its underlying sampling logic. To ensure a fair comparison with earlier models, we explicitly disable the reasoning capabilities of GPT5 by setting \texttt{reasoning\_effort = none}. All models are restricted to a \texttt{max\_tokens = 2048}. For each task, we generate a single candidate solution and post-process LLMs' responses to extract code blocks by discarding conversational text.

\section{Experiments}

\subsection{RQ1: How effective are LLMs for code generation in functional programming languages?}

To address this research question, following~\cite{lcb2024, 10.1145/3643674}, we evaluate performance via \texttt{pass@1}, i.e., the proportion of tasks where the model's first generated solution satisfies all test cases, and classify unsuccessful generations into: (1) \textit{Compilation Errors} (syntax/type mismatches); (2) \textit{Test Failures} (logic errors); and (3) \textit{Timeouts} (execution limits).

\textbf{Main Results.} Table~\ref{tab:pass_rate} presents the performance of the studied LLMs across four languages: Haskell, OCaml, Scala, and Java. Overall, we observe that LLM performance on functional programming improves substantially with model advancement, with GPT-5 achieving an approximately 3x increase in the number of functionally-correct code over GPT-3.5. However, a consistent performance gap persists between purely functional languages (Haskell and OCaml) and hybrid or imperative languages (Scala and Java).

Specifically, GPT-3.5 shows limited capability in functional languages, achieving pass rates of only 14.5\% in Haskell and 9.43\% in OCaml. In contrast, the model performs better in Scala (19.28\%) and Java (22.19\%). While GPT-4o and GPT5 substantially improves upon pass rate in Haskell (27.18\%/42.34\%) and Scala (36.2\%/52.16\%), it continues to exhibit a notable performance gap between the languages to Java. Specifically, GPT-4o’s pass rates in Haskell and OCaml lag those of Java by approximately 16.5 and 7.5 percentage points, respectively. This divergence even intensifies in GPT-5, where the performance gap widens to nearly 18 and 9 percentage points. These results suggest that despite the scaling and continuous improvement of Large Language Models, the performance gap between functional and imperative code generation remains significant.

\textbf{Compilation Errors.} Interestingly, we found that, different from imperative code generation, LLMs frequently generate uncompliable code with syntax or type errors in functional programming language. For instance, GPT-3.5 encounters high compilation error rates of 43.14\% and 25.5\% in OCaml and Haskell. Although GPT-4o manages to reduce these rates to approximately 21\%, it does not match the stability it demonstrates in Java (6.24\%). Notably, while GPT-5 successfully lowers the compilation error rate in OCaml to 13.73\% and achieves a remarkably low 4.62\% in Java, it fails to show similar improvement in Haskell, where the error rate remains high at 24.28\%. These findings indicate that even the most advanced models struggle to follow the strict syntactic rules of certain functional languages, likely due to the limited amount of such code in their training data compared to Java and other imperative languages.

\begin{table*}[ht!]
\centering
\begin{tabular}{llcccc}
\toprule
\textbf{Model} & \textbf{Metric} & \textbf{Haskell} & \textbf{OCaml} & \textbf{Scala} & \textbf{Java} \\
\midrule
\multirow{4}{*}{\textbf{GPT-3.5}} & Pass & 14.15\% & 9.43\% & 19.28\% & 22.19\% \\
 & Test Failures & 50.21\% & 43.13\% & 65.46\% & 57.28\% \\
 & Compilation Errors & 25.52\% & 43.14\% & 14.01\% & 10.40\% \\
 & Timeout & 10.12\% & 4.30\% & 1.25\% & 10.12\% \\
\midrule
\multirow{4}{*}{\textbf{GPT-4o}} & Pass & 27.18\% & 36.20\% & 38.83\% & 43.69\% \\
 & Test Failures & 45.08\% & 39.11\% & 51.60\% & 43.00\% \\
 & Compilation Errors & 21.36\% & 21.08\% & 9.57\% & 6.24\% \\
 & Timeout & 6.38\% & 3.61\% & 0\% & 7.07\% \\
\midrule
\multirow{4}{*}{\textbf{GPT-5}} & Pass & 42.34\% & 52.16\% & 58.36\% & 61.14\% \\
 & Test Failures & 32.63\% & 33.38\% & 32.37\% & 27.48\% \\
 & Compilation Errors & 24.28\% & 13.73\% & 9.27\% & 4.62\% \\
 & Timeout & 0.76\% & 0.72\% & 0\% & 6.76\% \\
\bottomrule
\end{tabular}
\caption{Effectiveness of studied LLMs for code generation in Haskell, Ocalm, Scala and Java.}
\label{tab:pass_rate}
\end{table*}

\subsection{RQ2: What are common code style and maintainability issues in code generated by LLMs in functional programming languages?}

To address this research question, we utilized a suite of language-specific static analysis tools to evaluate the adherence of LLM-generated code to best practices for code style and maintainability in functional programming. Specifically, we used HLint and GHC warnings for Haskell, dune and ocamlformat for Ocaml, Checkstyle and PMD for Java, and Scalastyle for Scala. We define a code snippet as ``clean'' only if it satisfies all applicable checks without any violations.

\textbf{Main Results.} Table~\ref{tab:clean-code-ratio} illustrates the proportion of ``clean'' code samples across both functionally-correct code generated by studied LLMs, i.e., those that passed all functional test cases in the first RQ.

For OCaml, while GPT-3.5 and GPT-4o maintain relatively high clean code ratios in these correct solutions (63\% and 61\%, respectively), GPT-5 shows a marked decline to 44\%. This suggests that while the older models fail more often, they tend to adhere to functional programming idioms and best practices in maintainability when they do succeed. In contrast, GPT-5, which achieves significantly higher correctness rates (as shown in Table~\ref{tab:pass_rate}), appears to prioritize functional correctness over these essential non-functional properties, potentially resorting to more imperative or complex structures to satisfy the test cases.

For Haskell, all studied LLMs consistently struggle to produce clean code. Specifically, GPT-5 and GPT-4o only achieve a clean code ratio of approximately 45-46\% of their functionally-correct solutions. Togeother with high compilation rates in RQ1, this consistently poor quality of LLM-generated code indicates that the strict constraints of Haskell remain a persistent challenge for current LLMs. Even the latest LLM, i.e., GPT-5, fails to improve code style and maintainability in Haskell. While this results are understandable given the complexity and strictliness of Haskell, it also highlight the limitations of LLMs in functional programming

Interestingly, our broader analysis reveals an unexpected trade-off in GPT-5 between code correctness and code qualiy, which happens not only in Ocalm and Haskell but also Scala and Java. Specifically, in RQ1, we can see that GPT-5 consistently improve the pass@1 accross four studied programming languages. However, in this experiment, we found that GPT-5 consistently witness lower code style and maintainability in all these languages compared to GPT-4o. Specifically, the clean code rate drop from 77\% $\to$ 57\% in Scala, 63\% $\to$ 51\% in Java, 46\% $\to$ 45\% in Haskell, 61\% $\to$ 44\% in Ocaml. This trend implies that as models optimize for higher pass rates, they may sacrifice code style and maintainability. This behaviors are consistent to "reward hacking", which are typically observed in LLMs. In practice, these behaviors can potentially harms software systems by introducing unoticed technical debt, which can pass functional test validations but increase the cost of maintainability. 

\begin{table}[t]
\centering
\begin{tabular}{lcccc}
\toprule
\multirow{2}{*}{\textbf{Model}} & \multicolumn{4}{c}{\textbf{Language}} \\
\cmidrule(lr){2-5}
& \textbf{Haskell} & \textbf{OCaml} & \textbf{Scala} & \textbf{Java} \\
\midrule
\textbf{GPT-3.5} & 22\% & 63\% & 61\% & 58\% \\
\textbf{GPT-4o}  & 46\% & 61\% & 77\% & 63\% \\
\textbf{GPT-5}   & 45\% & 44\% & 57\% & 51\% \\
\bottomrule
\end{tabular}
\caption{Proportion of clean code in functionally correct outputs generated by studied LLMs}
\label{tab:clean-code-ratio}
\vspace{-3mm}
\end{table}

\textbf{Common Issues.} To shed the light to future improvements, we conducted a semi-automated analysis on the low quality code with respects to best practices in functional programming detected by static analysis tools. We first manually identified recurring code style and maintainability issues and mapped them to specific diagnostics from our static analysis tools; subsequently, we developed a script to automatically parse the tool logs across the entire dataset to quantify the prevalence of these issues.

In Haskell, the most common issues were non-idiomatic patterns such as unnecessary lambdas (37×), missed eta reductions (36×), and redundant brackets. GHC further warned about unused variables and non-exhaustive pattern matches. In OCaml, most issues were code readability, raised by ocamlformat, e.g. redundant semicolons and parentheses. We also observed several critical issues, including type errors and unintended mutation of immutable data structures, although these occurred less frequently. Overall, while these issues do not compromise the functional correctness of LLM-generated code, they violate established best practices in functional programming and may increase the risk of future maintenance issues.

The most interesting issues are observed in Scala. Specifically, we found that 
the most frequent issue was the use of \texttt{return} (236x), which is discouraged in idiomatic Scala, as it indicate the use of imperative paradigm. 
Other violations included magic numbers, long methods, and inconsistent procedure definitions. The consistent presence of imperative constructs suggests that LLMs may not fully leverage Scala’s functional features, instead generating code that resembles Java-style imperative logic, which is allowed but not encouraged in Scala. These observations lead us to a broader question: \textit{Do LLMs truly generate functional code, or they just "hack" reward by embedding imperative code to achieve higher functional correctness?}

\textbf{Imperative Bias in LLMs-Generated Functional Code.} Although LLMs are explicitly queried to generate code in functional programming languages, we observed that their outputs often reflect an \textit{imperative programming bias}. This phenomenon is observed across all three languages, but is especially noticeable in Scala, with the use of constructs like \texttt{return}, mutable variables, and other Java-like patterns. In Haskell and OCaml, we observe similar biases, such as the use of mutable references, in-place updates, or IO effects in pure function features that conflict with the declarative and immutable nature of these languages. These observations suggest that LLMs do not fully adhere to expectations of functional programming. Rather, they tend to generalize from the dominant imperative structures in their training corpus, resulting in outputs that, while syntactically correct, often betray the core principles of FP: purity, immutability, and declarative control flow (see Appendix~\ref{appendix:imperative_patterns} for representative examples).

To validate the generality of these observations, we identified common imperative coding patterns, which are discouraged in functional programming principles. Particularly, we identified imperative and mutable constructs using keyword-based pattern matching (e.g., mutable collections, \texttt{var}, explicit loops in Scala; manual \texttt{let}-bindings, excessive \texttt{if-then-else} in Haskell; mutable references and assignments in OCaml). Then, we conducted a comprehensive analysis on LLM-generated code in Scala, Haskell, and OCaml across GPT-3.5, GPT-4o and GPT-5 to identify code with these imperative patterns. Table~\ref{tab:imperative_violations} presents the results of our analysis. We can see that all studied LLMs show a significant imperative bias in their generated code across all studied functional programming languages. Notably, GPT-5 demonstrates the highest prevalence of imperative patterns, with 94\%, 88\%, and 80\% of the generated code in Scala, Haskell, and OCaml, respectively, containing such constructs. These proportions represent a substantial increase relative to predecessors, i.e., GPT-3.5 and GPT-4o. These results suggest a ``reward hacking'' situation: while GPT-5 seem to become more capable of solving complex problems (as evidenced by the higher pass rates discussed previously), they increasingly rely on imperative shortcuts to ensure functional correctness rather than writing pure functional programming code.

\begin{table}[t]
\centering
\caption{Percentage of LLM-generated code with imperative patterns}
\begin{tabular}{lccc}
\toprule
\textbf{Language} & \textbf{GPT-3.5} & \textbf{GPT-4o}  & \textbf{GPT-5}\\
\midrule
Scala & 94\% & 88\% & 94\%\\
Haskell & 53\% & 63\% & 88\%\\
OCaml & 57\% & 42\% & 80\%\\
\bottomrule
\end{tabular}
\label{tab:imperative_violations}
\vspace{-3mm}
\end{table}

\subsection{RQ3: How effective are self-repair mechanisms in improving the correctness and code style and maintainability of LLM-generated functional code?}

Prior work~\cite{10.1145/3643674} suggests that LLMs can self-repair their mistakes with a feedback loop, this research question investigates whether LLM's \textit{self-repair ability} can improve the functional correctness and code style and maintainability of LLM outputs in functional programming languages. To this end, we examine the ability of ChatGPT to refine its own code given explicit repair prompts.

Following~\cite{10.1145/3643674}, we consider two following repair approaches: (1)\texttt{Simple Repair.} In this approach, we provide LLMs with a high-level system instruction to correct its previously generated solution. The instruction is formulated as in Appendix ~\ref{code:prompt_self_repair}. This baseline captures a minimal repair setting, where the model is asked to regenerate a better solution without any structured guidance beyond the task and functional programming reminder. (2)\texttt{Instruction-guided Repair.} In this apporach, we adopt a more structured approach with feedbacks from static analysis and hand-crafted instructions for common issues. Particularly, we ultised static analysis and test validations to identify potential function correctness and code style and maintainability issues. Then we break down the common quality issues observed in the generated code, such as compilation errors, type errors, imperative coding style, and provide the model with explicit instructions on how to fix them for each type of common issues. These instructions are tailored to each functional programming language (Haskell, OCaml, and Scala) and directly map common error categories to actionable repair instruction. These instructions are provided in ~\ref{tab:haskell_fix}, ~\ref{tab:ocaml_fix}, ~\ref{tab:scala_fix} in Appendix ~\ref{appendix:Prompting}.

\begin{table*}[t]
\centering
\caption{Number of correct solutions after applying different repair strategies across languages and models.}
\label{tab:self_repair}
\resizebox{\textwidth}{!}{
\begin{tabular}{l|ccc|ccc|ccc}
\toprule
 & \multicolumn{3}{c|}{GPT-3.5} & \multicolumn{3}{c|}{GPT-4o} & \multicolumn{3}{c}{GPT-5}\\
\textbf{Setting} & Haskell & OCaml & Scala & Haskell & OCaml & Scala & Haskell & OCaml & Scala\\
\midrule
Code Generated & 93 & 68 & 118 & 196 & 261 & 280 & 287 & 376 & 420\\
Simple Repair & \textbf{102} & 78 & 117 & 204 & \textbf{272} & \textbf{297} & \textbf{301} & \textbf{396} & 431\\
Instruction Repair & 98 & \textbf{85} & \textbf{139} & \textbf{209} & 266 & 288 & 295 & 391 & \textbf{433}\\
\bottomrule
\end{tabular}
}
\end{table*}

\textbf{Results.}
Table~\ref{tab:self_repair} details the number of functionally-correct solutions across three settings: initial generation, simple repair, and instruction-guided repair. Overall, we can see that LLMs consistently improve the functional correctness of their across all models and languages with self-repair capabilities using both studied repair approaches. A closer examination reveals a nuanced interaction between repair approaches and solution quality. While \texttt{Instruction-guided Repair} significantly boosts performance in certain contexts (e.g., GPT-3.5 in Scala: 118 $\to$ 139), it does not always strictly outperform \texttt{Simple Repair} in terms of raw pass rates. For instance, in Haskell and OCaml with GPT-5, \texttt{Simple Repair} achieves slightly higher correctness (e.g., Haskell: 301 vs. 295). This phenomenon is explained by analyzing code style and maintainability of the generated code. As in Table~\ref{tab:self_repair_clean}, in most of the cases, \texttt{Instruction-guided Repair} yields the highest rate of clean code, i.e., those follows best functional programming practices, substantially outperforming \texttt{Simple Repair}. This suggests that LLMs face a trade-off between code style and maintainability and functional correctness. \texttt{Simple Repair} allows the models to prioritize passing test cases by any necessary approaches, even retaining or introducing imperative shortcuts to fix errors. Meanwhile, \texttt{Instruction-guided Repair}, by enforcing strict FP constraints, increases the difficulty of the generation task. The models attempt to find a solution that is \textit{both} correct and high quality with respects to code style and maintainability. In cases where the model cannot synthesize a high quality solution, it may fail to generate valid code, resulting in a slight drop in pass rates despite the better code style and maintainability of the successful solutions. 

\begin{table}[t]
\centering
\caption{Proportion of clean code correct outputs generated by studied LLMs. Origin, Sim. Rep., and Ins. Rep. denote results obtained from the original generated code, code repaired using \texttt{Simple Repair}, and code repaired using \texttt{Instruction-guided Repair}, respectively.}
\label{tab:self_repair_clean}
\resizebox{\columnwidth}{!}{
\begin{tabular}{@{} l l rrr @{}}
\toprule
\textbf{Model} & \textbf{Method} & \textbf{Haskell} & \textbf{OCaml} & \textbf{Scala} \\ \midrule
\multirow{3}{*}{GPT-3.5} & Origin     & 22\% & 63\% & 61\% \\
                         & Sim. Rep.    & 46\% & 78\% & 75\% \\
                         & Ins. Rep. & \textbf{61\%} & \textbf{87}\% & \textbf{79\%} \\ \midrule
\multirow{3}{*}{GPT-4o}  & Origin     & 46\% & 61\% & \textbf{77\%} \\
                         & Sim. Rep.    & 48\% & \textbf{92\%} & 61\%  \\
                         & Ins. Rep. & \textbf{56\%} & 78\% & 66\% \\ \midrule
\multirow{3}{*}{GPT-5}   & Origin     & 45\% & 44\% &  57\%\\
                         & Sim. Rep.    & 58\% & 87\% & \textbf{62\%} \\
                         & Ins. Rep. & \textbf{63\%} & \textbf{91\%} &  60\%\\ \bottomrule
\end{tabular}
}
\end{table}

Fortunately the performance gap between \texttt{Instruction-guided Repair} and \texttt{Simple Repair} are minimal. For example, \texttt{Simple Repair} with GPT-5 only generate more 6 correct solutions than \texttt{Instruction-guided Repair} for Haskell. These results suggest that \texttt{Instruction-guided Repair} provides higher quality code with minimal trade-offs in \texttt{pass@1} rate. They also indicate that LLMs are increasingly capable of achieving functional correctness while simultaneously adhering to code style and maintainability considerations if best practices in functional programming are explicitly provided to these models. These findings suggest that  practices should be more explicitly incorporated into LLM training and inference to enable the generation of code that is both functionally correct and of high quality in terms of code style and maintainability.

\section{Related Works.}

\textbf{Code Generation Benchmarks.} Numerous datasets have been proposed to evaluate the code generation capabilities of Large Language Models (LLMs). Early benchmarks primarily focused on Python code generation, such as HumanEval~\cite{chen2021codex} and MBPP~\cite{austin2021program}. More recently, the research community has pivoted toward multilingual evaluation through datasets such as MultiPL-E~\cite{cassano2023multipl} and HumanEval-XL~\cite{peng2024humaneval}, though these still largely prioritize imperative programming paradigms (e.g., C++, Java, and Go). Other lines of work have examined specification-oriented languages, such as F*~\cite{Lahiri2024EvaluatingLU}, Dafny~\cite{Loughridge2024DafnyBenchAB} and JML~\cite{lecong2025llms}.
Different from these existing efforts, FPEval specifically targets functional programming languages, a paradigm that remains significantly under-represented in current benchmarking literature. Furthermore, we go beyond traditional functional correctness evaluation by conducting a holistic assessment of code style and maintainability. This holistic approach allows us to unveil deeper insights into the overall quality and technical debt of LLM-generated code in functional programming.

\textbf{Automation in Functional Programming.} Several works in the literature have also investigated the automation of functional programming. Notably, Gissurarson et al. proposed PropR \cite{gissurarson2022propr}, a property-based automated program repair (APR) framework designed to fix Haskell bugs by leveraging property-based testing and type-driven synthesis. Recently, Van Dam et al. \cite{van2024investigating} fine-tuned UniXCoder and CodeGPT specifically for Haskell code completion, demonstrating that language-specific models can significantly outperform general-purpose LLMs in this domain. Different from these studies, which are primarily restricted to the Haskell ecosystem, our work expands the scope to a diverse set of functional programming languages, including Haskell, Scala, and OCaml. Furthermore, rather than focusing on fine-tuning smaller architectures, we conduct a comprehensive evaluation of the latest frontier LLMs, including GPT-3.5, GPT-4o, and GPT-5. This allow us to assess the capabilities of these state-of-the-art models across multiple functional languages, providing a broader perspective on the current boundaries of AI-driven software engineering in functional programming.

\section{Conclusion}

In this paper, we presented FPEval, the first comprehensive evaluation framework and empirical study for assessing the code generation abilities of Large Language Models within the functional programming paradigm. FPEval contains an dual-assessment approach, containing both rigorous test suites and static analysis tools, allowed us to not only evaluate functional correctness but also code style and maintainability of LLM-generated code.

Our findings reveal a significant "paradigm gap" in current LLM performance: these models demonstrate substantially higher error rates in pure functional programming languages, such as Haskell and OCaml, compared to hybrid or imperative baselines. Furthermore, our analysis indicates that even when LLMs achieve functional correctness, they frequently generate low quality code that violate best practice in functional programming, thereby undermining the safety and maintainability benefits inherent to functional programming. 

Despite these limitations, we also found that the code generation capabilities of LLMs in functional programming improves with model advancement. Additionally, our results demonstrated that LLMs hold potential on self-repairing their mistakes with guided by detailed feedbacks from static analysis and customized instructions. Overall, we hope that FPEval can serve as robust benchmark for the community and offer empirical insights to guide the development of LLMs in functional programming.

\section{Limitations}

The limitations of this work are as follows:

First, our evaluation is constrained the limited scope of the subject models. Due to computational and financial constraints, our evaluation focuses exclusively on three Large Language Models (LLMs): GPT-3.5, GPT-4o, and GPT-5. This narrow selection may impact the generalizability of our findings across the broader landscape of open-source or alternative proprietary models. To minimize this risk, we have taken specific steps: (1) we selected models from OpenAI, the current industry leader, to ensure our results reflect the state-of-the-art capabilities in the field. (2) we selected models across three distinct generations of OpenAI GPT series to facilitate a longitudinal analysis of LLM evolution and performance trajectories. (3) we included GPT-5, the most advanced model available at the time of experimentation, to ensure that our benchmarks capture the upper bound of current LLMs abilities.

Second, this study mainly focuses on the code generation capabilities of LLMs in functional programming. While the generation of functionally-correct and high-quality code is a fundamental requirement for practical software engineering, it represents only a subset of the developer's workflow. Our evaluation does not account for other essential software development tasks, such as bug repair or code optimization. Consequently, our findings reflect the models' effectiveness in initial synthesis but may not generalize to their performance in iterative software evolution or debugging. However, we hope that our work can serve as a foundational benchmark for functional programming in the LLM era, providing a necessary baseline upon which future studies into broader software engineering automation can be built.

Finally, our benchmark primarily comprises problems sourced from educational and competitive programming platforms. While these tasks provide rigorous constraints for evaluating algorithmic correctness, they may not be fully representative of industrial-scale functional programming. Real-world software involves complex inter-module dependencies, legacy architectural constraints, and specific domain-driven design patterns that are often absent in isolated programming challenges. In the future work, we will consider the expansion of our dataset to include large-scale, open-source functional programming repositories. 

\bibliography{main}

@misc{openai2024gpt4technicalreport,
      title={GPT-4 Technical Report}, 
      author={OpenAI},
      year={2023},
      eprint={2303.08774},
      archivePrefix={arXiv},
      primaryClass={cs.CL},
      url={https://arxiv.org/abs/2303.08774}, 
}

@inproceedings{
    lcb2024,
    title={LiveCodeBench: Holistic and Contamination Free Evaluation of Large Language Models for Code},
    author={Naman Jain and King Han and Alex Gu and Wen-Ding Li and Fanjia Yan and Tianjun Zhang and Sida Wang and Armando Solar-Lezama and Koushik Sen and Ion Stoica},
    booktitle={The Thirteenth International Conference on Learning Representations},
    year={2025},
    url={https://openreview.net/forum?id=chfJJYC3iL}
}

@misc{cursor2023,
  title={Cursor: The {AI}-first Code Editor},
  author={Anysphere},
  year={2023},
  url={https://cursor.sh},
  note={Accessed: 2025-11-19}
}

@misc{anthropic2025claude,
  title={Claude {Code}: An Agentic Coding Tool},
  author={Anthropic},
  year={2025},
  url={https://www.anthropic.com/news/claude-code},
  note={Technical Release}
}

@article{2022AlphaCode,
  author = {Yujia Li and David Choi and Junyoung Chung and Nate Kushman and Julian Schrittwieser and R\'{e}mi Leblond and Tom Eccles and James Keeling and Felix Gimeno and Agustin Dal Lago and Thomas Hubert and Peter Choy and Cyprien de Masson d'Autume and Igor Babuschkin and Xinyun Chen and Po-Sen Huang and Johannes Welbl and Sven Gowal and Alexey Cherepanov and James Molloy and Daniel J. Mankowitz and Esme Sutherland Robson and Pushmeet Kohli and Nando de Freitas and Koray Kavukcuoglu and Oriol Vinyals},
  title = {Competition-level code generation with AlphaCode},
  journal = {Science},
  volume = {378},
  number = {6624},
  pages = {1092-1097},
  year = {2022},
  doi = {10.1126/science.abq1158},
  URL = {https://www.science.org/doi/abs/10.1126/science.abq1158},
  eprint = {https://www.science.org/doi/pdf/10.1126/science.abq1158}
}

@INPROCEEDINGS{10172803,
  author={Xia, Chunqiu Steven and Wei, Yuxiang and Zhang, Lingming},
  booktitle={2023 IEEE/ACM 45th International Conference on Software Engineering (ICSE)}, 
  title={Automated Program Repair in the Era of Large Pre-trained Language Models}, 
  year={2023},
  volume={},
  number={},
  pages={1482-1494},
  doi={10.1109/ICSE48619.2023.00129}}

@INPROCEEDINGS{Fan2023LLMSurvey,
  author={Fan, Angela and Gokkaya, Beliz and Harman, Mark and Lyubarskiy, Mitya and Sengupta, Shubho and Yoo, Shin and Zhang, Jie M.},
  booktitle={2023 IEEE/ACM International Conference on Software Engineering: Future of Software Engineering (ICSE-FoSE)}, 
  title={Large Language Models for Software Engineering: Survey and Open Problems}, 
  year={2023},
  volume={},
  number={},
  pages={31-53},
  doi={10.1109/ICSE-FoSE59343.2023.00008}}

@misc{mitchell2024hlint,
  author = {Neil Mitchell},
  title = {{HLint}: Haskell Source Code Suggestions},
  year = {2024},
  publisher = {GitHub},
  howpublished = {\url{https://github.com/ndmitchell/hlint}},
  note = {Accessed: 2024-05-20}
}

@misc{ocamlformat2024,
  author = {{OCamlFormat Developers}},
  title = {{OCamlFormat}: Auto-formatter for OCaml Code},
  year = {2024},
  publisher = {GitHub},
  howpublished = {\url{https://github.com/ocaml-ppx/ocamlformat}},
  note = {Accessed: 2024-05-20}
}

@misc{scalastyle2024,
  author = {Matthew Farwell and {Scalastyle Contributors}},
  title = {{Scalastyle}: Scala Style Checker},
  year = {2024},
  howpublished = {\url{http://www.scalastyle.org}},
  note = {Accessed: 2024-05-20}
}

@article{Hughes1989WhyFP,
  title={Why Functional Programming Matters},
  author={John Hughes},
  journal={Comput. J.},
  year={1989},
  volume={32},
  pages={98-107},
}

@inproceedings{Achten2013WhyFP,
  title={Why Functional Programming Matters to Me},
  author={Peter Achten},
  booktitle={The Beauty of Functional Code},
  year={2013},
}

@article{Ray2014ALS,
  title={A large scale study of programming languages and code quality in github},
  author={Baishakhi Ray and Daryl Posnett and Vladimir Filkov and Premkumar T. Devanbu},
  journal={Proceedings of the 22nd ACM SIGSOFT International Symposium on Foundations of Software Engineering},
  year={2014},
}

@article{Xu2023AreWR,
  title={Are We Ready to Embrace Generative AI for Software Q\&A?},
  author={Bowen Xu and Thanh-Dat Nguyen and Thanh Le-Cong and Thong Hoang and Jiakun Liu and Kisub Kim and Chen Gong and Changan Niu and Chenyu Wang and Bach Le and David Lo},
  journal={2023 38th IEEE/ACM International Conference on Automated Software Engineering (ASE)},
  year={2023},
  pages={1713-1717},
}

@ARTICLE{TechnicalDebt,
  author={Kruchten, Philippe and Nord, Robert L. and Ozkaya, Ipek},
  journal={IEEE Software}, 
  title={Technical Debt: From Metaphor to Theory and Practice}, 
  year={2012},
  volume={29},
  number={6},
  pages={18-21},
  doi={10.1109/MS.2012.167}}

@inproceedings{Perry_2023, series={CCS ’23},
   title={Do Users Write More Insecure Code with AI Assistants?},
   url={http://dx.doi.org/10.1145/3576915.3623157},
   DOI={10.1145/3576915.3623157},
   booktitle={Proceedings of the 2023 ACM SIGSAC Conference on Computer and Communications Security},
   publisher={ACM},
   author={Perry, Neil and Srivastava, Megha and Kumar, Deepak and Boneh, Dan},
   year={2023},
   month=nov, pages={2785–2799},
   collection={CCS ’23} }

@article{10.1145/3643674,
author = {Liu, Yue and Le-Cong, Thanh and Widyasari, Ratnadira and Tantithamthavorn, Chakkrit and Li, Li and Le, Xuan-Bach D. and Lo, David},
title = {Refining ChatGPT-Generated Code: Characterizing and Mitigating Code Quality Issues},
year = {2024},
issue_date = {June 2024},
publisher = {Association for Computing Machinery},
address = {New York, NY, USA},
volume = {33},
number = {5},
issn = {1049-331X},
url = {https://doi.org/10.1145/3643674},
doi = {10.1145/3643674},
journal = {ACM Trans. Softw. Eng. Methodol.},
month = jun,
articleno = {116},
numpages = {26}
}

@article{Prather2023ItsWT,
  title={“It’s Weird That it Knows What I Want”: Usability and Interactions with Copilot for Novice Programmers},
  author={James Prather and Brent N. Reeves and Paul Denny and Brett A. Becker and Juho Leinonen and Andrew Luxton-Reilly and Garrett B. Powell and James Finnie-Ansley and Eddie Antonio Santos},
  journal={ACM Transactions on Computer-Human Interaction},
  year={2023},
  volume={31},
  pages={1 - 31},
}

@article{Kazemitabaar2023StudyingTE,
  title={Studying the effect of AI Code Generators on Supporting Novice Learners in Introductory Programming},
  author={Majeed Kazemitabaar and Justin T.H. Chow and Carl Ka To Ma and Barb Ericson and David Weintrop and Tovi Grossman},
  journal={Proceedings of the 2023 CHI Conference on Human Factors in Computing Systems},
  year={2023},
}

@misc{OpenAI2022ChatGPT,
  author = {OpenAI},
  title = {Introducing ChatGPT},
  year = {2022},
  month = {11},
  day = {30},
  url = {https://openai.com/index/chatgpt/},
  urldate = {2025-12-29},
  note = {Accessed: 2025-12-29}
}

@misc{OpenAI2025GPT5,
  author = {OpenAI},
  title = {Introducing GPT-5},
  year = {2025},
  month = {8},
  day = {7},
  url = {https://openai.com/index/introducing-gpt-5/},
  urldate = {2025-12-29},
  note = {Accessed: 2025-12-29}
}

@inproceedings{van2024investigating,
  title={Investigating the performance of language models for completing code in functional programming languages: a haskell case study},
  author={Van Dam, Tim and Van der Heijden, Frank and De Bekker, Philippe and Nieuwschepen, Berend and Otten, Marc and Izadi, Maliheh},
  booktitle={Proceedings of the 2024 IEEE/ACM First International Conference on AI Foundation Models and Software Engineering},
  pages={91--102},
  year={2024}
}

@inproceedings{gissurarson2022propr,
  title={Propr: property-based automatic program repair},
  author={Gissurarson, Matth{\'\i}as P{\'a}ll and Applis, Leonhard and Panichella, Annibale and Van Deursen, Arie and Sands, David},
  booktitle={Proceedings of the 44th international conference on software engineering},
  pages={1768--1780},
  year={2022}
}

@inproceedings{peng2024humaneval,
  title={HumanEval-XL: A Multilingual Code Generation Benchmark for Cross-lingual Natural Language Generalization},
  author={Peng, Qiwei and Chai, Yekun and Li, Xuhong},
  booktitle={Proceedings of the 2024 Joint International Conference on Computational Linguistics, Language Resources and Evaluation (LREC-COLING 2024)},
  pages={8383--8394},
  year={2024}
}

@article{cassano2023multipl,
  title={Multipl-e: A scalable and polyglot approach to benchmarking neural code generation},
  author={Cassano, Federico and Gouwar, John and Nguyen, Daniel and Nguyen, Sydney and Phipps-Costin, Luna and Pinckney, Donald and Yee, Ming-Ho and Zi, Yangtian and Anderson, Carolyn Jane and Feldman, Molly Q and others},
  journal={IEEE Transactions on Software Engineering},
  volume={49},
  number={7},
  pages={3675--3691},
  year={2023},
  publisher={IEEE}
}

@article{austin2021program,
  title={Program synthesis with large language models},
  author={Austin, Jacob and Odena, Augustus and Nye, Maxwell and Bosma, Maarten and Michalewski, Henryk and Dohan, David and Jiang, Ellen and Cai, Carrie and Terry, Michael and Le, Quoc and others},
  journal={arXiv preprint arXiv:2108.07732},
  year={2021}
}

@article{chen2021codex,
  title={Evaluating Large Language Models Trained on Code},
  author={Mark Chen and Jerry Tworek and Heewoo Jun and Qiming Yuan and Henrique Ponde de Oliveira Pinto and Jared Kaplan and Harri Edwards and Yuri Burda and Nicholas Joseph and Greg Brockman and Alex Ray and Raul Puri and Gretchen Krueger and Michael Petrov and Heidy Khlaaf and Girish Sastry and Pamela Mishkin and Brooke Chan and Scott Gray and Nick Ryder and Mikhail Pavlov and Alethea Power and Lukasz Kaiser and Mohammad Bavarian and Clemens Winter and Philippe Tillet and Felipe Petroski Such and Dave Cummings and Matthias Plappert and Fotios Chantzis and Elizabeth Barnes and Ariel Herbert-Voss and William Hebgen Guss and Alex Nichol and Alex Paino and Nikolas Tezak and Jie Tang and Igor Babuschkin and Suchir Balaji and Shantanu Jain and William Saunders and Christopher Hesse and Andrew N. Carr and Jan Leike and Josh Achiam and Vedant Misra and Evan Morikawa and Alec Radford and Matthew Knight and Miles Brundage and Mira Murati and Katie Mayer and Peter Welinder and Bob McGrew and Dario Amodei and Sam McCandlish and Ilya Sutskever and Wojciech Zaremba},
  year={2021},
  eprint={2107.03374},
  archivePrefix={arXiv},
  primaryClass={cs.LG}
}

@misc{LeCong2026MemoryEfficient,
      title={Memory-Efficient Large Language Models for Program Repair with Semantic-Guided Patch Generation}, 
      author={Thanh Le-Cong and Bach Le and Toby Murray},
      year={2025},
      eprint={2410.16655},
      archivePrefix={arXiv},
      primaryClass={cs.SE},
      url={https://arxiv.org/abs/2410.16655}, 
}

@misc{leetcode,
  title        = {LeetCode: The World's Leading Online Programming Learning Platform},
  author       = {{LeetCode, Inc.}},
  year         = {2026},
  url          = {https://leetcode.com/},
  note         = {Accessed: 2026-01-05},
  organization = {LeetCode, LLC},
  description  = {An online platform for practicing coding problems and preparing for software engineering interviews. Founded in 2015 in Silicon Valley, LeetCode provides algorithmic exercises, contests, and discussion forums used globally by programmers and job seekers.}  
}

@inproceedings{lecong2025llms,
  author    = {Le-Cong, Thanh and Le, Bach and Murray, Toby},
  title     = {Can LLMs Reason About Program Semantics? A Comprehensive Evaluation of LLMs on Formal Specification Inference},
  booktitle = {Proceedings of the 63rd Annual Meeting of the Association for Computational Linguistics (Volume 1: Long Papers)},
  year      = {2025},
  address   = {Vienna, Austria},
  publisher = {Association for Computational Linguistics},
  pages     = {21991--22014},
  doi       = {10.18653/v1/2025.acl-long.1068},
  url       = {https://aclanthology.org/2025.acl-long.1068/}
}

@article{Lahiri2024EvaluatingLU,
  title={Evaluating LLM-driven User-Intent Formalization for Verification-Aware Languages},
  author={Shuvendu K. Lahiri},
  journal={2024 Formal Methods in Computer-Aided Design (FMCAD)},
  year={2024},
  pages={142-147},
}

@article{Loughridge2024DafnyBenchAB,
  title={DafnyBench: A Benchmark for Formal Software Verification},
  author={Chloe Loughridge and Qinyi Sun and Seth Ahrenbach and Federico Cassano and Chuyue Sun and Ying Sheng and Anish Mudide and Md Rakib Hossain Misu and Nada Amin and Max Tegmark},
  journal={Trans. Mach. Learn. Res.},
  year={2024},
  volume={2025},
  url={https://api.semanticscholar.org/CorpusID:270391562}
}

\appendix
\section{Prompt for Synthesizing Test-Case Generators}
~\label{appendix:prompt_testcases}
To automatically construct high-quality and diverse test cases for each programming task, we employ GPT-4o to synthesize Python input-generation code directly from the natural-language task description. The model is prompted to behave as an expert competitive programmer and to strictly follow all problem constraints when constructing test data.
Below is the exact prompt template used for generating test-case synthesis code.
The assistant receives (i) the task description and (ii) a Python template scaffold, and is instructed to output a single Python function named construct\_inputs() that returns 10 diverse inputs without printing them.
\begin{lstlisting}[language=Python, breaklines=true, basicstyle=\ttfamily\footnotesize, title={Prompt for input generation}, label={code:input_gen_prompt}]
    ("system",
     "You are an expert Python competitive programmer and your goal is to construct input generators for testing programming contest problems. You will write relevant generators and finally construct `construct_inputs` function that returns a list of diverse inputs sampled from the generator. Remember to strictly follow the instructions and constraints present in the problem statement.\n"
    ),
    ("human",
     "### INSTRUCTIONS:\n"
     "- Identify the constraints from the problem statement.\n"
     "- Construct a Python function `construct_inputs()` that generates diverse test cases.\n"
     "- Ensure all generated values strictly follow the constraints.\n\n"
     "### EXAMPLE REFERENCE:\n"
     "import numpy as np\n"
     "def random_input_generator(weight_min, weight_max, size_min, size_max):\n"
     "    weights_size = np.random.randint(size_min, size_max+1)\n"
     "    weights = np.random.randint(weight_min, weight_max, size=weights_size).tolist()\n"
     "    k = np.random.randint(1, len(weights)+1)\n"
     "    return weights, k\n\n"
     "def construct_inputs():\n"
     "    inputs_list = []\n"
     "    ## random inputs\n"
     "    for i in range(8):\n"
     "        inputs_list.append(random_input_generator(1, 50, 1, 10))  # Adjust according to constraints\n"
     "    return inputs_list\n\n"
     "### TASK:\n"
     "{task_description}\n"
     "{python_template}\n"
     "Construct a random input generator. Use the format used in the above example by returning a single function that builds diverse inputs named `construct_inputs`. Do not print the inputs_list.\n"),
\end{lstlisting}
\section{Test Execution.}~\label{appendix:test_execution}  

To ensure the execution of public and private test cases,  we developed a specialized test infrastructure for functional languages that operates in four stages. First, FPEval generates per-problem unit tests using standard libraries (e.g., \texttt{OUnit2} for OCaml, \texttt{HUnit} for Haskell). Second, it constructs compilation-ready test templates, into which the LLM-generated solutions are injected. Third, the infrastructure validates both public and private test cases by compiling and executing the resulting files in an isolated environment using Docker. Finally, it captures execution logs to rigorously distinguish between syntax errors, static type mismatches, runtime failures and functional errors. The infrastructure also automatically handles differences in data formats (e.g., booleans, lists, strings), supports docstring-based type extraction, and performs type-aware formatting of test inputs and expected outputs. This automation was essential for scaling evaluation across hundreds of problems and ensuring a high-quality assessment of LLM effectiveness in functional programming languages.

\section{Data Contamination.}

\begin{figure}[t]
\centering
\includegraphics[width=\linewidth]{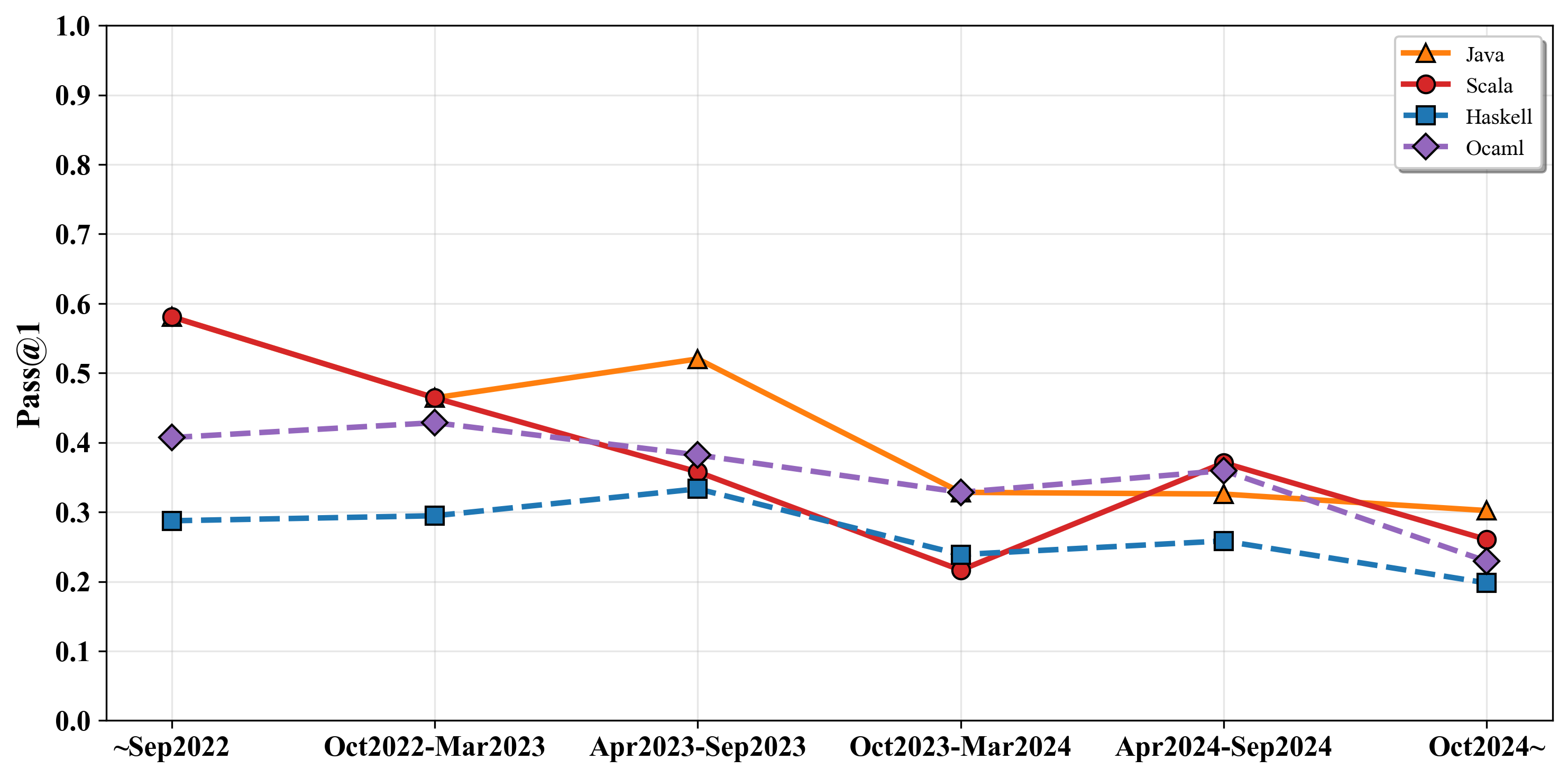}
\caption{Pass@1 performance of GPT-4o}
\label{fig:cutoff_gpt4o}
\end{figure}

We also investigate the potential impact of data contamination of studied LLMs. We specifically select GPT-4o for this analysis due to its knowledge cutoff, which provides a balanced distribution of tasks between the pre-cutoff and post-cutoff periods in our evaluation date. To quantify the impact of data contamination, we partition the programming tasks into six distinct chronological periods based on their release dates:(1) before September 2022, (2) October 2022–March 2023, (3) April–September 2023, (4) October 2023–March 2024, (5) April–September 2024, and (6) after October 2024. Figure ~\ref{fig:cutoff_gpt4o} illustrates the pass rates of GPT-4o in four studied programming languages. We can see that GPT-4o show show a substantial post-cutoff performance drop in Java and Scala, i.e., from 52.0\% to 32.8\% in Java and from 35.8\% to 21.6\% in Scala. This results suggest possible data contamination in pre-training data. In contrast, the model maintains a low and stable accuracy in Ocaml and Haskell throughout, consistent with their limited presence in pre-training corpora.

\section{Static Analysis Rule Categorization}
\label{appendix:static_analysis_rules}

To ensure the reproducibility of our code cleanliness metrics, we provide a categorization of the static analysis rules employed by \texttt{FPEval}. Table~\ref{tab:static_analysis_rules} details the specific configurations used for each language. These rules are selected to detect two primary types of violations:
\begin{itemize}
    \item \textbf{Imperative Bias:} Constructs that are syntactically valid but violate functional programming principles (e.g., use of mutable variables, explicit loops, or procedural control flow).
    \item \textbf{Non-Idiomatic Usage:} Code that fails to leverage the expressive power of the language.
\end{itemize}

For \textbf{Haskell}, we rely on \texttt{HLint} to detect redundant constructs and suggest higher-order replacements. For \textbf{OCaml}, we invoke the compiler directly using the \texttt{-warn-error +a} flag to enforce strict static semantics. For \textbf{Scala}, we employ the \textbf{default} \texttt{Scalastyle} configuration, which reflects the community's standard for idiomatic code. However, to prioritize semantic correctness over formatting minutiae, we explicitly disabled all whitespace-related rules (e.g., tab checking, spacing alignment), focusing strictly on logic and paradigm violations.

\begin{table*}[ht!]
\centering
\caption{Categorization of Static Analysis Rules employed in FPEval}
\label{tab:static_analysis_rules}
\resizebox{\textwidth}{!}{%
\begin{tabular}{l l l p{8cm}}
\toprule
\textbf{Language} & \textbf{Tool} & \textbf{Rule / ID} & \textbf{Description \& Rationale} \\
\midrule
\multirow{5}{*}{\textbf{Haskell}} 
 & \multirow{4}{*}{HLint} 
 & \textit{Use map / fmap} & Suggests replacing explicit recursion or list comprehensions with higher-order functions (e.g., \texttt{map}), enforcing idiomatic FP style. \\
 & & \textit{Redundant do} & Flags \texttt{do} blocks containing a single expression, encouraging pure functional expressions over monadic sequencing. \\
 & & \textit{Use let} & Recommends \texttt{let} bindings over monadic generators (\texttt{<-}) for pure values. \\
 & & \textit{Eta reduce} & Encourages point-free style by removing redundant function arguments. \\
 \cmidrule(lr){2-4}
 & GHC & \textit{-Wall} & Enables all warnings, strictly flagging shadowed variables and unused matches common in imperative logic porting. \\
\midrule
\multirow{5}{*}{\textbf{Scala}} 
 & \multirow{5}{*}{\makecell[l]{Scalastyle\\(Default Profile)}} 
 & \textit{VarChecker} & \textit{(Standard)} Detects mutable variables (\texttt{var}), promoting the use of immutable values (\texttt{val}). \\
 & & \textit{ReturnChecker} & \textit{(Standard)} Detects the explicit \texttt{return} keyword, ensuring expression-oriented control flow. \\
 & & \textit{NullChecker} & \textit{(Standard)} Disallows \texttt{null}, enforcing type-safe error handling via \texttt{Option}. \\
 & & \textit{CyclomaticComplexity} & \textit{(Standard)} Penalizes methods with high complexity ($>10$), often caused by nested imperative \texttt{if-else} blocks. \\
 & & \textit{PublicMethodsHaveType} & \textit{(Standard)} Requires explicit return types for public methods, ensuring type clarity in functional APIs. \\
\midrule
\multirow{4}{*}{\textbf{OCaml}} 
 & \multirow{4}{*}{\makecell[l]{Compiler\\(\texttt{ocaml -c})}} 
 & \textit{-warn-error +a} & \textbf{Strict Mode:} Treats all warnings as fatal errors. This ensures generated code adheres strictly to OCaml static semantics. \\
 & & \textit{Warning 8} & \textit{Partial match:} Enforced via strict mode. Rejects pattern matching that fails to cover all cases. \\
 & & \textit{Warning 26} & \textit{Unused variable:} Enforced via strict mode. Rejects unused bindings often left behind by procedural logic generation. \\
 \cmidrule(lr){2-4}
 & OCamlFormat & \textit{Formatting} & Enforces community-standard indentation, discouraging deep nesting structures typical of imperative code. \\
\bottomrule
\end{tabular}
}
\end{table*}

\section{Representative Examples of Imperative Code Patterns}
\label{appendix:imperative_patterns}
\paragraph{1. Mutable Variables}

\begin{itemize}
    \item \textbf{Imperative-style (incorrect):}
    \begin{lstlisting}[language=Scala]
var count = 0
count = count + 1
  
    
// Functional-style (preferred)
   
val count = 0
val newCount = count + 1
    \end{lstlisting}
    
    In FP, variables are ideally immutable (\texttt{val} in Scala, \texttt{let} without reassignment in OCaml). However, GPT often generates code using mutable bindings (\texttt{var}) or reassignment operators.
\end{itemize}

\paragraph{2. Imperative Loops}

\begin{itemize}
    \item \textbf{Imperative-style (incorrect):}
    \begin{lstlisting}[language=Scala]
for (i <- 0 until n) {
  println(i)
}
    \end{lstlisting}

    \item \textbf{Functional-style (preferred):}
    \begin{lstlisting}[language=Scala]
(0 until n).foreach(println)
    \end{lstlisting}

    \item \textbf{Or recursion (in OCaml):}
    \begin{lstlisting}[language=OCaml]
let rec print_numbers i n =
  if i < n then (
    print_int i;
    print_newline ();
    print_numbers (i + 1) n
  )
    \end{lstlisting}

    GPT frequently defaults to \texttt{for}, \texttt{while}, or \texttt{do while} loops common in imperative languages but less idiomatic in FP. A functional solution would use higher-order functions or recursion.
\end{itemize}

\paragraph{3. In-place Mutation of Data Structures}

\begin{itemize}
    \item \textbf{Imperative-style (incorrect):}
    \begin{lstlisting}[language=Scala]
val arr = Array(1, 2, 3)
arr(0) = 42
    \end{lstlisting}

    \item \textbf{Functional-style (preferred):}
    \begin{lstlisting}[language=Scala]
val arr = Vector(1, 2, 3)
val newArr = arr.updated(0, 42)
    \end{lstlisting}

    Instead of treating data structures as immutable, GPT often produces code that mutates arrays or lists directly, especially in performance-related contexts. This contradicts the functional principle of constructing new values rather than mutating existing ones.
\end{itemize}

\paragraph{4. Use of \texttt{null} Instead of Option Types}

\begin{itemize}
    \item \textbf{Imperative-style (incorrect):}
    \begin{lstlisting}[language=Scala]
def findUser(name: String): User = {
  if (db.contains(name)) db(name) else null
}
    \end{lstlisting}

    \item \textbf{Functional-style (preferred):}
    \begin{lstlisting}[language=Scala]
def findUser(name: String): Option[User] = db.get(name)
    \end{lstlisting}

    FP discourages the use of \texttt{null}, favoring safer alternatives like \texttt{Option}, \texttt{Maybe}, or pattern matching.
\end{itemize}

\section{Imperative Bias in LLM-Generated Functional Code}
\label{appendix:imperative_patterns_LLM}

Although LLMs are prompted to generate code in functional programming languages such as Haskell, OCaml, and Scala, their outputs often reflect an \textit{imperative programming bias}. This section provides representative examples and quantitative evidence of this phenomenon.
\paragraph{Example 1: Scala}
\begin{itemize}
    \item \textbf{Imperative-style (incorrect):}
    \begin{lstlisting}[language=Scala, 
    title ={Problem: Count alternating subarrays},
    captionpos=b,]
import scala.collection.mutable

def countAlternatingSubarrays(nums: List[Int]): Int = {
    val n = nums.length                         
    // Mutable variable 'n'
    var count = 0                              
    //Mutable variable (var) 
    var left = 0                                
    //Mutable variable (var)
    var right = 0                               
    // Mutable variable (var)
    val set = mutable.HashSet[Int]()            
    // Uses mutable collection (HashSet)
    while (right < n) {                         
    //Imperative loop
        set.add(nums(right))                    
        // Side-effect: mutating set
        while (set.size > 2) {
            set.remove(nums(left))              
            //Side-effect: mutating set
            left += 1                           
            //Mutation
        }
        count += right - left + 1               
        //Mutation
        right += 1                              
        //Mutation
    }

    count
}

    \end{lstlisting}

    \item \textbf{Functional-style (preferred):}
    \begin{lstlisting}[language=Scala]
def countAlternatingSubarrays(nums: List[Int]): Int = {
  nums.indices.foldLeft((0, 0, Map.empty[Int, Int], 0)) {
    case ((left, right, freqMap, count), r) =>
      val num = nums(r)
      val updatedFreq = freqMap.updated(num, freqMap.getOrElse(num, 0) + 1)

      // Slide left pointer while more than 2 distinct elements
      val (newLeft, newFreq) = Iterator.iterate((left, updatedFreq)) {
        case (l, fmap) =>
          val leftNum = nums(l)
          val newCount = fmap(leftNum) - 1
          val newFmap = if (newCount == 0) fmap - leftNum else fmap.updated(leftNum, newCount)
          (l + 1, newFmap)
      }.dropWhile { case (_, fmap) => fmap.size > 2 }.next()

      val newCount = count + (r - newLeft + 1)
      (newLeft, r + 1, newFreq, newCount)
  }._4
}
    \end{lstlisting}

The functional version uses an immutable `Map` instead of a mutable `HashSet`, eliminates all `var` declarations, and replaces reassignment with state threading via `foldLeft`. It also employs `Iterator.iterate` to express the sliding window logic in a recursion-inspired style.

\end{itemize}

\paragraph{Example 2: OCaml}
\begin{itemize}
    \item \textbf{Imperative-style (incorrect):}
    \begin{lstlisting}[language=ocaml, 
    title ={Problem: Find the minimum amount of time to brew potions},
    captionpos=b,]
let minTime (skill: int list) (mana: int list) : int =
  let n = List.length skill in
  let m = List.length mana in
  let dp = Array.make_matrix n m 0 in  
  (*uses mutable array*)
  dp.(0).(0) <- skill.(0) * mana.(0);  
  (*mutation using "<-"*)
  for j = 1 to m - 1 do               
  (*Imperative loop *)
    dp.(0).(j) <- max dp.(0).(j-1) (skill.(0) * mana.(j));  
    (*Mutation inside loop*)
  done;
  for i = 1 to n - 1 do               
  (*Imperative loop*)
    dp.(i).(0) <- dp.(i-1).(0) + skill.(i) * mana.(0);       
    (*Mutation*)
    for j = 1 to m - 1 do             
    (*Nested loop*)
      dp.(i).(j) <- max dp.(i).(j-1) dp.(i-1).(j) + skill.(i) * mana.(j); 
      (*Mutation*)
    done;
  done;
  dp.(n-1).(m-1)               

    \end{lstlisting}

    \item \textbf{Functional-style (preferred):}
    \begin{lstlisting}[language=ocaml]
let minTime (skill: int list) (mana: int list) : int =
  let rec build_dp prev_row i =
    match List.nth_opt skill i with
    | None -> prev_row
    | Some s ->
      let row =
        List.mapi (fun j m ->
          let from_left = if j = 0 then None else Some (List.nth row (j - 1)) in
          let from_top = List.nth prev_row j in
          let cell_value =
            match from_left with
            | None -> from_top + s * m
            | Some l -> max l from_top + s * m
          in
          cell_value
        ) mana
      in
      build_dp row (i + 1)
  in
  match skill with
  | [] -> 0
  | s0 :: rest ->
    let first_row =
      snd (List.fold_left (fun (max_val, acc) m ->
        let curr = max max_val (s0 * m) in
        (curr, acc @ [curr])
      ) (min_int, []) mana)
    in
    let final_dp = build_dp first_row 1 in
    List.hd (List.rev final_dp)

    \end{lstlisting}
\end{itemize}
In this version:
\begin{itemize}
    \item The dp matrix is reconstructed immutably using List.fold\_left and pure functions.
    \item There are no imperative loops or mutation (<- is gone).
    \item Recursion and list transformations maintain purity and immutability.
\end{itemize}

\paragraph{Example 3: Haskell}
\begin{itemize}
    \item \textbf{Imperative-style (incorrect):}
    \begin{lstlisting}[language=haskell, 
    title ={Problem: Maximum number of operations with the same score ii},
    captionpos=b,]
import Data.List
maxOperations :: [Int] -> Int
maxOperations nums = helper nums 0
  where
    helper nums count
      | length nums < 2 = count
      | otherwise = let score = sum $ take 2 nums
            in helper (delete (head nums) $ delete (head $ reverse nums) nums) (count + 1)       

    \end{lstlisting}

    \item \textbf{Functional-style (preferred):}
    \begin{lstlisting}[language=haskell]
maxOperations = go 0
  where
    go count (x:xs@(_:_)) =
      let mid = init xs  
      in go (count + 1) mid 
    go count _ = count

    \end{lstlisting}
\end{itemize}

\section{Prompt Design and Language-specific Self-repair Instructions}
\label{appendix:Prompting}
This appendix details the prompting strategies and language-specific repair mappings used in our experiments. It consists of two parts: (1) the prompt templates for code generation and self-repair, and (2) the mapping between common compiler warnings or errors and the corresponding self-repair instructions for each functional language (Haskell, OCaml, and Scala).

\subsection{Prompt Templates}

The first part presents the exact prompts used for the baseline code generation and self-repair settings.

\begin{lstlisting}[ title={Prompt for generating code}, label={code:prompt}]
    ("system",
     "You are an expert {lang} programmer. You will be given a question (problem specification) and will generate a correct {lang} program that matches the specification and passes all tests. You will NOT return anything except for the program AND necessary imports.\n",
    ),
    ("human",
     "### QUESTION:\n{description}\n"
     "### FORMAT: You will use the following starter code to write the solution to the problem and enclose your code within delimiters.\n{template}\n"
     "### ANSWER: (use the provided format with backticks)\n"
     )

\end{lstlisting}

\begin{lstlisting}[ title={Prompt Baseline for self-repair}, label={code:prompt_self_repair}]
    ("system",
    "You are a helpful programming assistant and an expert {lang} programmer. The previously generated code has quality issues, does not pass the tests, and is not idiomatic functional programming. Please provide a better code implementation as expected by the task description and the function using idiomatic functional programming style. You must put the entire fixed program within code delimiters only once.",
    ),
    ("human",
    "### CODE: \n{pre_code}\n"
    )
\end{lstlisting}
\begin{quote}
\texttt{}
\end{quote}
\subsection{Language-specific Self-repair Mappings}

The second part summarizes detailed mappings between compiler feedback and the corresponding self-repair instructions. These mappings serve as structured guidance for the model to correct errors and improve code style and maintainability during the self-repair process.

\begin{table*}[ht!]
\centering
\caption{Instruction mapping for Haskell self-repair}
\label{tab:haskell_fix}
\begin{tabular}{p{0.3\linewidth} p{0.65\linewidth}}
\toprule
\textbf{Warning / Error} & \textbf{Instruction for Self-repair} \\
\midrule
Avoid lambda & Avoid using lambda expressions when a function or operator already exists. \\
Eta reduce & Remove unnecessary parameters by applying eta-reduction. \\
Redundant bracket & Eliminate redundant brackets to simplify the expression. \\
Avoid lambda using infix & Replace lambda expressions with infix operators when possible. \\
Use zipWith & Use \texttt{zipWith} for parallel list operations instead of combining \texttt{map} and \texttt{zip}. \\
Avoid reverse & Avoid using \texttt{reverse} before \texttt{head}, \texttt{last}, or indexing, as it is inefficient. \\
Move brackets to avoid \$ & Prefer parentheses over excessive use of the \texttt{\$} operator for readability. \\
Move filter & Apply \texttt{filter} closer to the data source to reduce unnecessary computation. \\
Use uncurry & Use \texttt{uncurry} when applying a function to a tuple. \\
Use infix & Use infix notation for better readability when working with operators. \\
\bottomrule
\end{tabular}
\end{table*}

\begin{table*}[ht!]
\centering
\caption{Instruction mapping for OCaml self-repair}
\label{tab:ocaml_fix}
\begin{tabular}{p{0.3\linewidth} p{0.65\linewidth}}
\toprule
\textbf{Warning / Error} & \textbf{Instruction for Self-repair} \\
\midrule
Type error & Check type annotations and ensure expressions match expected types. For example, if a function expects an \texttt{int} but receives a \texttt{string}, convert or adjust the type. \\
Unbound identifier & Declare the identifier before using it, or fix the spelling. Make sure the correct module is opened/imported. \\
Parse error & Fix OCaml syntax: check missing \texttt{in}, \texttt{->}, \texttt{;;}, or misplaced parentheses/brackets. \\
Unused & Remove unused variables, or prefix them with \texttt{\_} if they are intentionally unused. \\
Exhaustiveness error & Add missing cases in pattern matching to cover all constructors, or use \texttt{\_} as a catch-all case. \\
Incorrect arity & Check the number of arguments when calling a function. Supply missing arguments or remove extra ones. \\
Missing labeled argument & Add the required labeled argument when calling the function (e.g., \texttt{f \textasciitilde x:value} if the function expects \texttt{\textasciitilde x}). \\
Mutation of immutable & OCaml values are immutable by default. Rewrite code to avoid mutation, and use new bindings (\texttt{let x = ...}) instead of trying to update old ones. \\
\bottomrule
\end{tabular}
\end{table*}

\begin{table*}[ht!]
\centering
\caption{Instruction mapping for Scala self-repair}
\label{tab:scala_fix}
\begin{tabular}{p{0.35\linewidth} p{0.6\linewidth}}
\toprule
\textbf{Warning / Error} & \textbf{Instruction for Self-repair} \\
\midrule
If block needs braces & Always wrap \texttt{if} blocks with braces \texttt{\{\}} for clarity and to avoid ambiguity. \\
Avoid using return & Do not use \texttt{return}; rely on expression values instead. \\
Magic Number & Replace magic numbers with named constants or enums for readability. \\
Cyclomatic complexity of & Refactor the function into smaller ones or use functional constructs (\texttt{map}, \texttt{fold}, recursion) to reduce complexity. \\
There should be a space before the plus (+) sign & Add proper spacing around operators (e.g., \texttt{a + b} not \texttt{a+b}). \\
File line length exceeds & Split long lines into multiple shorter lines for readability. \\
File must end with newline character & Ensure the file ends with a newline character. \\
\bottomrule
\end{tabular}
\end{table*}

\end{document}